\documentclass[aps,pre,twocolumn,showpacs,superscriptaddress,groupedaddress]{revtex4-1}
\usepackage{amsmath}
\usepackage{amssymb}
\usepackage{amsfonts}
\usepackage{graphicx}
\usepackage{dcolumn}
\usepackage{bm}
\usepackage{sidecap}
\usepackage{subfloat}
\usepackage{parskip}
\usepackage{grffile}
\usepackage{color}
\usepackage{hyperref}
\usepackage{hhline}
\usepackage{mathtools}
\usepackage{graphics}
\usepackage{multirow}
\usepackage{verbatim}
\usepackage{longtable}
\usepackage{rotating}
\usepackage{setspace}
\usepackage{epsfig}
\usepackage{subfigure}
\usepackage{epstopdf}
\usepackage{gensymb}
\usepackage[normalem]{ulem}
\usepackage[table]{xcolor}

\newcommand{\A}{\textbf{A}}
\newcommand{\I}{\textbf{I}}
\newcommand{\M}{\cal{M}}

\makeatletter
\def\@eqnnum{{\normalsize \normalcolor (\theequation)}}
 \makeatother
 
\graphicspath{{./}{ER/main/}}

\begin{document}
\title{Inhibition induced explosive synchronization in multiplex networks}
\author{Sarika Jalan$^{1,2}$}
\email{sarikajalan9@gmail.com} 
\author{Vasundhara Rathore$^2$, Ajay Deep Kachhvah$^1$ and Alok Yadav$^1$}
\affiliation{1. Complex Systems Lab, Discipline of Physics, Indian Institute of Technology Indore, Khandwa Road, Simrol, Indore-453552, India} 
\affiliation{2. Discipline of Biosciences and Biomedical Engineering, Indian Institute of Technology Indore, Khandwa Road, Simrol, Indore-453552, India}
\date{\today}


\begin{abstract}
To date, explosive synchronization (ES) is shown to be originated from either degree-frequency correlation or inertia of phase oscillators. Of late, it has been shown that ES can be induced in a network by adaptively controlled phase oscillators. Here we show that ES is a generic phenomenon and can occur in any network by appropriately multiplexing it with another layer. We devise an approach which leads to the occurrence of ES with hysteresis loop in a network upon its multiplexing with a negatively coupled (or inhibitory) layer. We discuss the impact of various structural properties of positively coupled (or excitatory) and inhibitory layer along with the strength of multiplexing in gaining control over the induced ES transition. This investigation is a step forward in highlighting the importance of multiplex framework not only in bringing novel phenomena which are not possible in an isolated network but also in providing more structural control over the induced phenomena.
\end{abstract}


 \maketitle
\section{Introduction}
Synchronization of networked phase oscillators has proven itself to be an important process in understanding the collective behavior of a variety of real-world complex systems ranging from physical to biological systems \cite{sync_appl1,sync_appl2}. In a recent study, Garde\~{n}es \textit{et al.} \cite{ES_first} reported that the transition to synchronization can be an abrupt or first-order type, called explosive synchronization (ES), and can be achieved by setting a correlation between the natural frequencies and respective degrees of networked phase oscillators. Owing to the significance of ES in elucidating various abrupt transitions found in real-world systems, for an instance, large black-outs (cascading failure of the power stations)~\cite{ES_power} or epileptic seizure (the abrupt synchronous firing of neurons)~\cite{ES_sez} and chronic pain in the FM brain~\cite{ES_pain}, ES has received tremendous attention from network science community~\cite{ES_work}. Tanaka {\it et al.}\cite{ES_inertia}, for the first time introduced the occurrence of the first-order (discontinuous) transition resulting from finite inertia in the networked Kuramoto oscillators. Of late, further studies on ES have demonstrated that the microscopic correlation between degree-frequency is not the only criteria for the occurrence of ES. For an instance, ES is also shown to be resulted from a fraction of adaptively controlled oscillators~\cite{ES_multi_f,ES_classicalkuramoto}, and by assuming a positive correlation between coupling strengths and respective absolute of natural frequencies of networked oscillators~\cite{ES_coup} in isolated networks. All the investigations reinforce the fact that any suppressing factor which hampers the merging of small synchronous clusters, lead to abrupt formation of single giant synchronous component~\cite{ES_rev}.
Lately, the investigations on ES have been extended to multilayered networks by considering a fraction of adaptively controlled Kuramoto oscillators~\cite{ES_multi_f}, second-order Kuramoto oscillators (with inertia)~\cite{ES_multi_s, ES_delay_mp}, and intertwined multilayer couplings~\cite{ES_multi_t}. A multiplex network~\cite{multi_def, mp_tune, mp_so} is a framework of interconnected layers, each with different connectivity explicating different dynamical processes, however, represented by a common set of nodes. It provides a more accurate representation of many real-world networks~\cite{multi_appl}.
Further, the inhibitory coupling is known to suppress synchronization in networked phase oscillators \cite{mp_ES_frust, chimera_rep}. For instance, inhibition is shown to be a significant factor in controlling excessive synchronization in neurons which is known to destroy complex interaction patterns in brain network and eventually lead to diseases like the epileptic seizure~\cite{inhi_appl2}. In a recent study~\cite{ES_neg}, ES is shown to be suppressed in an isolated network upon introduction of negatively coupled oscillators.
Hence, in the current study, we demonstrate that one can induce ES by multiplexing with negatively coupled oscillators. In fact, the ES can be induced in any given network, without employing degree-frequency correlation or any other type of structural correlation, by multiplexing it with a layer with appropriate connectivity. We multiplex a layer of positively coupled nodes to a layer with negatively coupled nodes bringing the suppressive effect in the system. We, further, conjecture that we can find a critical negative coupling strength, through multiplexing, which could produce suppressive effect sufficient enough to prevent the formation of the giant cluster, leading to ES. Eventually, we prove our conjecture to be true by demonstrating the occurrence of ES in the excitatory layer for a variety of multiplex networks. Further, we discuss the impact of various structural parameters on induced ES. \\
\begin{figure}[t]
 \centering
  \includegraphics[width=1\columnwidth]{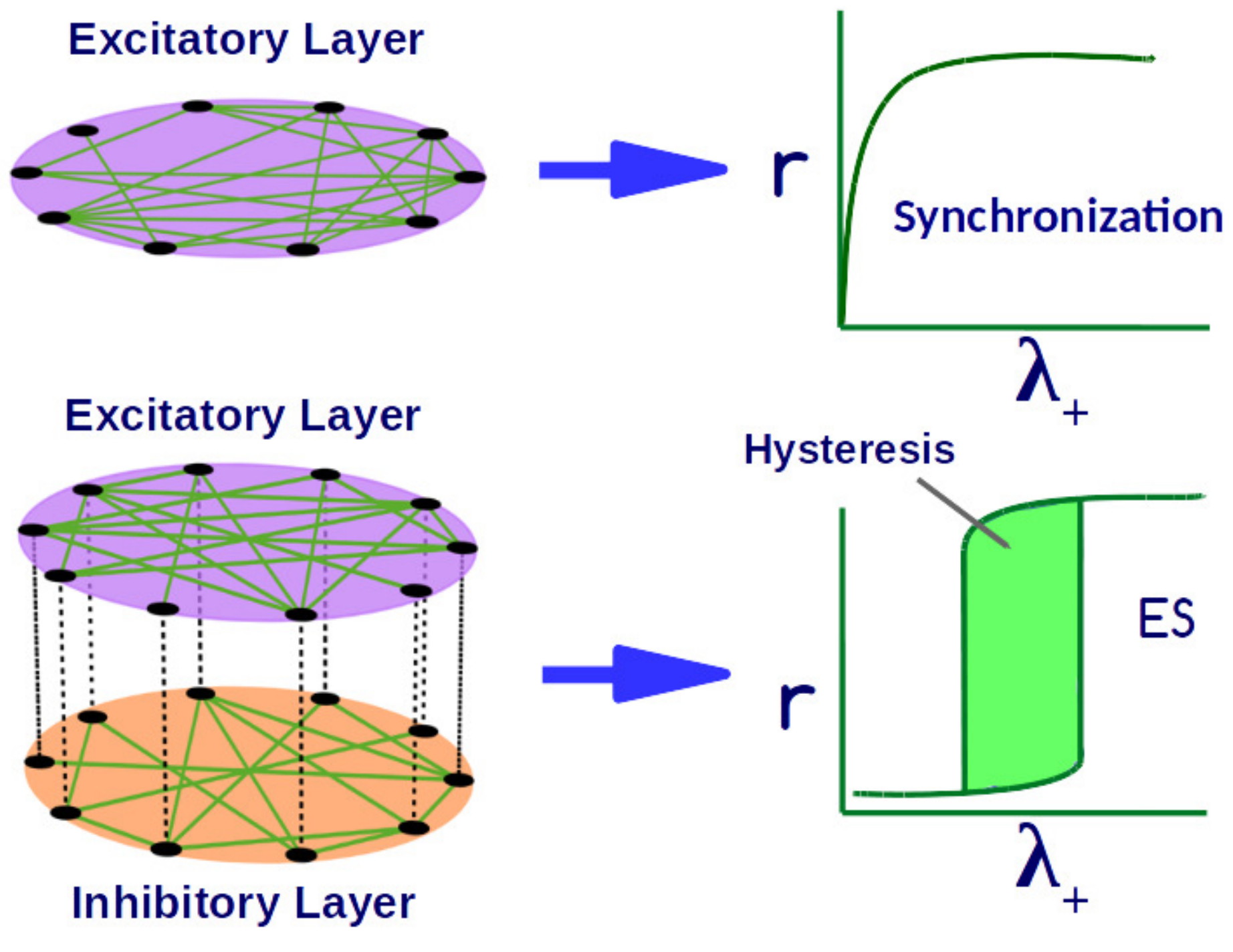}
   \caption{(Color Online) Schematic diagram for a single layer network where nodes are positively coupled depicting a smooth second order transition to synchrony (top). Multiplex network in which same nodes are positively and negatively coupled separately in two different layers, called excitatory and inhibitory, respectively (bottom). Parameter $r$ depict the degree of coherence as a function of coupling strength $\lambda$ as defined in Eq.\ref{eq:ordp_multi}. The positively coupled layer shows ES transition with hysteresis upon multiplexing with a negative layer.}
     \label{figure1}
\end{figure}
\section{Methods}
In the current work, we aim to explore how the path to synchronization in a network is affected by multiplexing it with an inhibitory layer of a multiplex network. To accomplish this, we consider an undirected and unweighted multiplex network having $N$ nodes in $M$ layers. Here, the dynamics of each node is determined by the most celebrated Kuramoto oscillators \cite{kuramoto_book_1}. To incorporate inhibition, one layer of the multiplex network is subject to the inhibitory coupling between the nodes. Here, we focus on investigating the effect of inhibition on intra-layer synchronization in multiplexed layers with different sets of choices for network topologies. For the sake of comparison, we will restrict our study to two-layered multiplex networks. The time evolution of Kuramoto oscillators in a duplex network with such scenario is governed by 
    \begin{align} \label{eq:k_model_multi}
    	\dot\theta^i_1 &= \omega^i_1+ \lambda_+ \sum_{j=1}^{N} A^{ij}_1 [\sin(\theta_1^j-\theta_1^i)]+ D_x [\sin(\theta_2^i-\theta_1^i)], \nonumber  \\
        \dot\theta^i_2 &= \omega^i_2+ \lambda_-\sum_{j=1}^{N} A^{ij}_2 [\sin(\theta_2^j-\theta_2^i)]+ D_x [\sin(\theta_1^i-\theta_2^i)], 
    \end{align}
    
\begin{figure}[t]
 \centering
   \includegraphics[width=1\columnwidth]{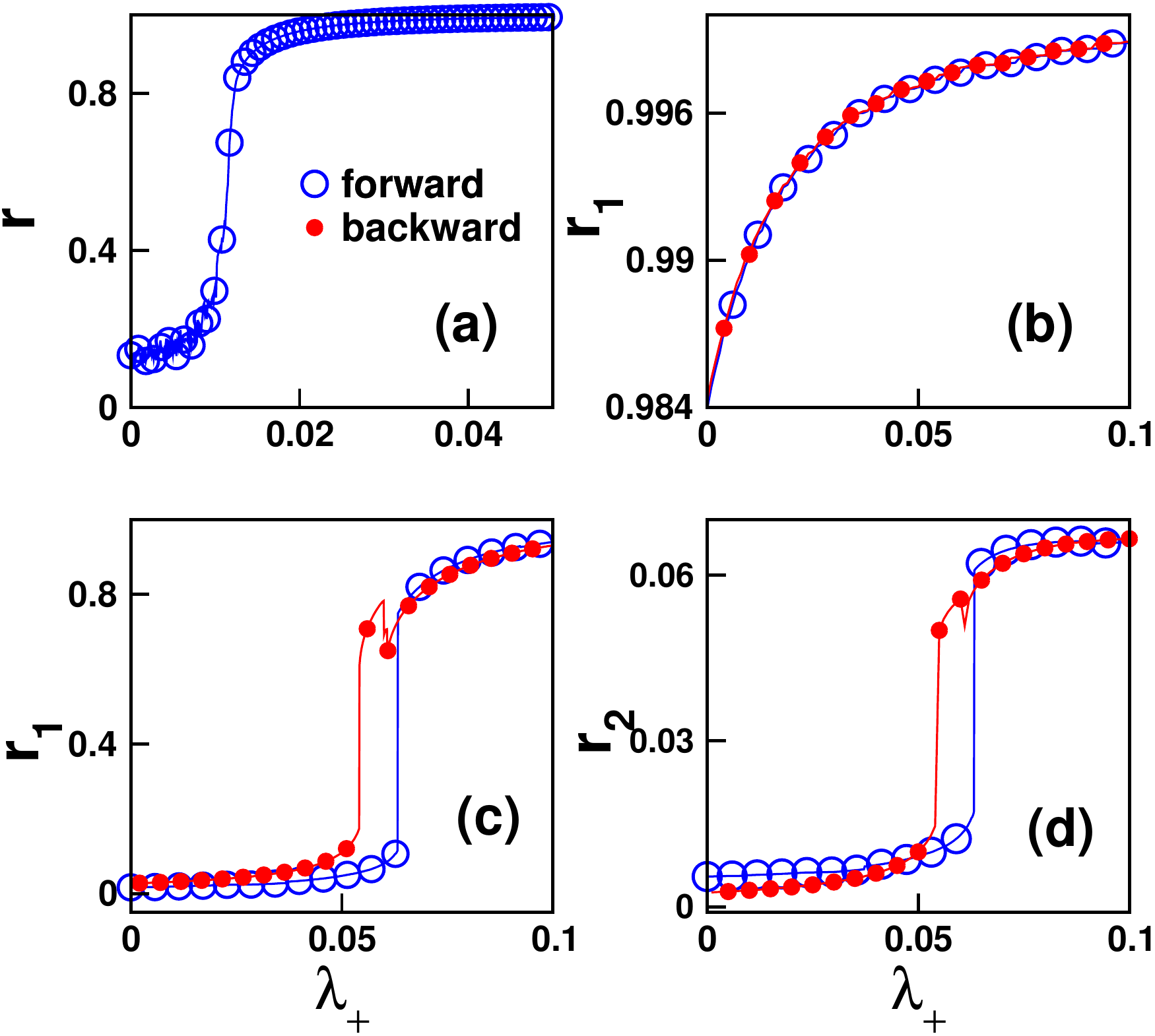}
   \caption{(Color Online) {Order parameters illustrating continuous transition, as a function of coupling strength, in a positively coupled GC network (a) in its isolation and (b) when it is multiplexed with another positively coupled regular layer. When the GC layer is instead multiplexed with a negatively coupled ($\lambda_{-}=2$) regular layer, (c) it exhibits ES transition with associated hysteresis loop, (d) while the transition in the inhibitory layer gets suppressed after exhibiting a very short abrupt jump.}} 
     \label{figure2}
      \end{figure}
where the subscripts $1$ and $2$ symbolize two distinct layers, $\theta^i(t)$ and $\omega^i$ with $i=1,...,N$ denote the initial phase and intrinsic frequency of the $i^{th}$ node, respectively. The parameter $D_x$ denote the inter-layer coupling strength between the layers. The positive coupling $(\lambda_+ > 0)$ reflects the positive interactions between the nodes of excitatory layer while negative coupling $(\lambda_- < 0)$ accounts for the negative interactions or suppressive behavior in the inhibitory layer.
The intra-layer connectivity between the nodes following a network topology is encoded in the adjacency matrix $A$ (of dimension $M \times N$) such that $A^{ij}=1$ $(0)$ if $i^{th}$ and $j^{th}$ nodes are connected (disconnected). Hence, the adjacency matrix of the multiplex network can be denoted by the set of intra-layer adjacencies, $\{\A_1,\A_2,..., \A_M \}$. For a duplex network, adjacency matrix is given by
\begin{align}
\M=\left(
\begin{array}{cc}
\A_1  & D_x\I \\ 
D_x\I & \A_2 \\
\end{array}
\right)
\end{align}
where $\I$ is the identity matrix. A schematic representation of a two-layered multiplex network with excitatory (positive) and inhibitory (negative) layers connected via inter-layer coupling strength $D_x$ is given in Fig.\ref{figure1}.
To track the degree of coherence or synchronization in the multiplex network, we define the global order parameters $r_1$ and $r_2$ for both the layers in terms of average phases $\phi_1$ and $\phi_2$ as
\begin{align} \label{eq:ordp_multi}
    r_1(t)e^{\imath \phi_1} &= \frac{1}{N}\sum_{k=1}^{N}e^{\imath\theta_1^k}, \nonumber \\
    r_2(t)e^{\imath \phi_2} &= \frac{1}{N}\sum_{k=1}^{N}e^{\imath\theta_2^k}.
\end{align}
Hence $r_1=1$ and $r_2=1$ for completely synchronous states for both the layers, and $r_1=0$ and $r_2=0$ would imply to total incoherent states.

Additionally, if $T$ be the total time (long enough) of averaging after discarding initial transients $t_r$ of the system states, the effective frequency $\langle{\omega^i}\rangle$ of each node in a network is defined as 
\begin{equation}
\langle{\omega^i}\rangle=\frac{1}{T}\int_{t_r}^{t_r+T}\dot\theta^i(t)\mathrm{d}t.
\end{equation}
We also define a symmetric matrix encoding the degree of coherence between every pair of linked nodes as~\cite{def_rlink}
\begin{equation}
r^{ij} =A^{ij}\left |\lim_{T\rightarrow\infty}\frac{1}{T}\int_{t_r}^{t_r+T}e^{\iota\left[\theta^i(t)-\theta^j(t)\right]}\mathrm{d}t\right|\;,
\label{eq:rij}
\end{equation}
so that $0\leq r^{ij}\leq1$. $r^{ij}=1$ if a pair $(i,j)$ of linked nodes are completely coherent and $r^{ij}=0$ if completely incoherent. Hence, the local composition of the synchronization patterns in a network can be captured by the fraction of all synchronized links defined as
\begin{equation}
r^{link}=\frac{1}{2N_c}\sum_{i}\sum_{j} r^{ij},
\label{r_link}
\end{equation}
where $2N_c$ is the total number of existing links in a network.

\section{Numerical Results}
In this section, we will discuss in details various numerical results exploring the behavior of the path to synchronization in the excitatory layer when it is multiplexed with an inhibitory layer. To achieve this, we study the synchronization profile of each layer by computing order parameters $r_1$ and $r_2$ defined in Eq.\ref{eq:ordp_multi} as a function of the coupling strength.

To induct inhibition between the coupling in the inhibitory layer, we fix $\lambda_-$ to a constant negative value. Further, to track transition to synchronization, the coupling strength $\lambda_+$ is varied. Firstly, $\lambda_+$ is increased adiabatically starting from $\lambda_+=0$ (incoherent state) to a $\lambda_+ + n\delta\lambda_+$ corresponding to a synchronous state, in the step of $\delta\lambda_+$. Secondly, to verify the existence of hysteresis, we adiabatically decrease $\lambda_+$ from $\lambda_+ + n\delta\lambda_+$ (synchronous) to $\lambda_+=0$ (incoherent) in the step of $\delta\lambda_+$. We name the above two processes as forward and backward transitions, respectively. The order parameters $r_1$ and $r_2$ are computed at each step $\delta\lambda_+$ during increment as well as decrement of $\lambda_+$.
For our simulations, we have taken $\delta\lambda_+=10^{-3}$. We integrate the system (Eq.\ref{eq:k_model_multi}) using the RK4 method with step size $dt=0.01$ for long enough time ($5\times10^4$ time steps) to arrive at a stationary state and eliminate initial transients. For both the layers, initial values of phases $\theta^i$ and natural frequencies $\omega^i$ are drawn uniformly randomly in the range $[0,2\pi)$ and $[-0.5,0.5]$, respectively. For the sake of comparison and to perform a variety of analysis, {we have considered a duplex network comprising a globally connected GC (excitatory) and regular ring (inhibitory) layers, each having $N=50$ nodes with $D_x=2$, otherwise mentioned elsewhere}.
\begin{figure*}
	\centering
	\includegraphics[height=8 cm,width=16 cm]{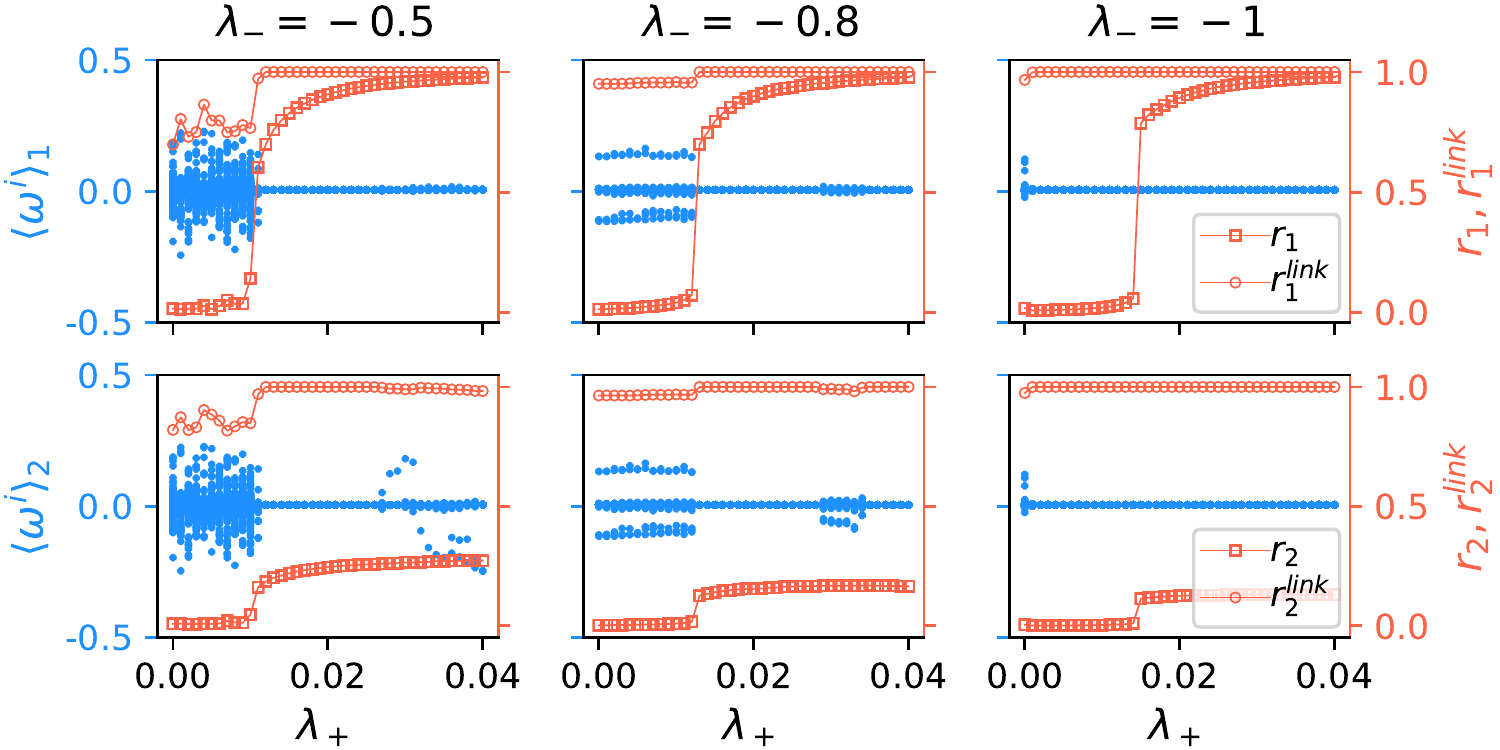}\\
	\caption{(Color Online) Effective frequencies, order parameters and $r^{link}$ of both the excitatory and inhibitory layers as a function of $\lambda_+$ for different strengths of inhibitory coupling $\lambda_{-}$.
	The number of nodes in each layer $N = 200$ with $D_x=2$ and average degree $\langle k_2 \rangle=10$ of the inhibitory layer.}
	\label{figure3}
\end{figure*}

\begin{figure}
	\centering
	\includegraphics[height=3 cm,width=8.5 cm]{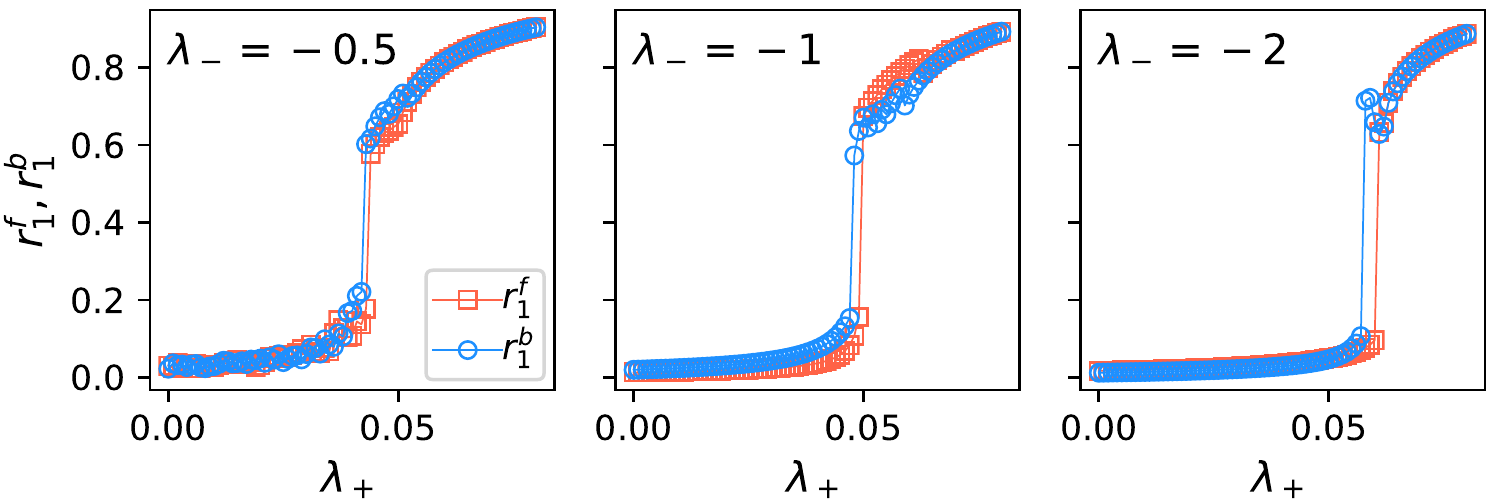}\\
	\caption{(Color Online) Order parameters $r_1^f$ and $r_1^b$ respectively for forward and backward transitions of the excitatory layer (of $N=50$ nodes with $D_x=2$) for different values of inhibitory coupling strength demonstrating hysteresis curves.}
	\label{figure3_n1}
\end{figure}

\begin{figure}
	\centering
	\includegraphics[height=4.5 cm,width=5.5 cm]{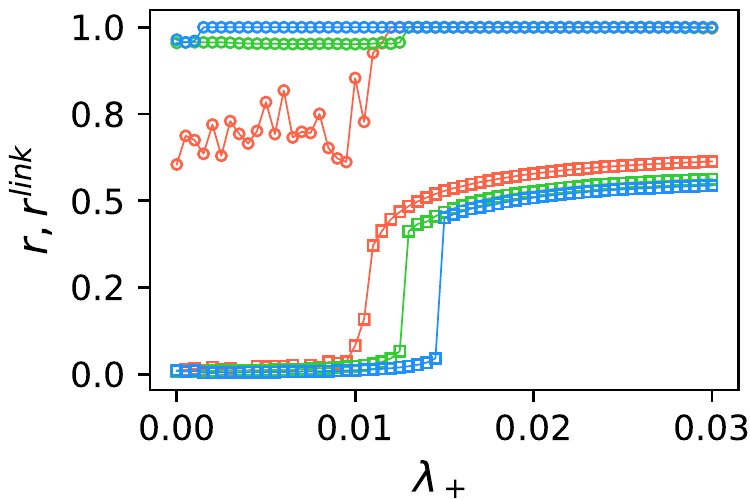}\\
	\caption{(Color Online) Order parameter $r$ (square) and $r^{link}$ (circle) of the entire network as a function of $\lambda_+$ for different values of $\lambda_-$. The red, green and blue colors correspond to $\lambda_-=-0.5, -0.8$  and $-1$, respectively. The results are included for $N=200$ nodes in each layer.}
	\label{figure3_n2}
\end{figure}

It is known that an isolated GC layer exhibits second order transition (Fig.\ref{figure2} (a)). When the GC layer is multiplexed with an positively coupled layer (regular ring network), the second order nature of transition in the GC layer still persists with no hysteresis observed in the forward and the backward transitions (Fig.\ref{figure2} (b)). 
Next, when the GC layer is multiplexed with a negatively coupled (inhibitory) layer {with a fixed $\lambda_-$ which is considerably larger than the critical value of $\lambda_+$ required for transition of the excitatory layer, strikingly} the GC layer ($r_1$) starts exhibiting ES (first order) transition with an associated hysteresis loop in the forward and the backward transitions (Fig.\ref{figure2} (c)).
Moreover, the value of order parameter $r_2$ for the inhibitory layer remains low ($r_2<0.1$), i.e., it does not show global synchronization \textbf{(see Fig.\ref{figure2} (d)}). This behavior of the inhibitory layer is similar to that of the isolated one. However, closely examining the value of $r_2$ reveals that the inhibitory layer too experiences the ES transition with a hysteresis loop at the same value of the forward and backward coupling strengths as those of the excitatory layer. However, the jump size is very small as it gets suppressed by the strong inhibitory coupling within the layer.

{\em \textbf{Behavior of $\langle\omega^i \rangle$, $r$ and $r^{link}$ with $\lambda_{-}$}}:
{To have a deeper understanding about the underlying micro-dynamics taking place in the asynchronous and synchronous states of both the layers, we closely look at behavior of microscopic properties such as effective frequencies and $r^{link}$ for each layer.}

{Fig.~\ref{figure3} illustrates the behavior of order parameters, effective frequencies and $r^{link}$ of both the layers for different values of inhibitory coupling strength. For a rather weak $\lambda_-=-0.5$, the effective frequencies of the excitatory and inhibitory nodes are spread over a considerable width until a critical value of $\lambda_+$ triggers the onset of ES. For any value of $\lambda_+$ prior to ES threshold, the values of $r^{link}_1$ and $r^{link}_2$ range between $0.7-0.85$ and $0.8-0.9$, respectively, while $r_1$ and $r_2$ tend to zero. It indicates that despite global incoherence there exists noticeable local clustering of the phases in both the layers. 
At the brink of transition, nevertheless, the excitatory layer exhibits ES overcoming the inhibitory force as the existing locally clustered phases abruptly construct the giant synchronized cluster. At the same time, $r_2$ gets suppressed sporting a very short ES jump due to the impression of unrelenting intralayer inhibition. $r^{link}_2=0.99$ and $\langle\omega^i\rangle_2\simeq0$ for any value of $\lambda_+$ post the transition, hence, the inhibitory layer maintains even stronger local clustering of phases.
For a bit stronger $\lambda_-=-0.8$, the width of the spread of $\langle\omega^i\rangle_1$ and $\langle\omega^i\rangle_2$ shrinks and they abruptly converge to mean frequency at the outset of ES at relatively higher $\lambda_+$ yielding a more profound ES jump than that for $\lambda_-=-0.5$ case. For any value of $\lambda_+$ prior to ES threshold, higher $r^{link}_1=0.95$ and $r^{link}_2=0.96$ imply the existence of more significant local clustering patterns than that for $\lambda_-=-0.5$ case. 
A strong enough $\lambda_-=-1$ leads to the oscillation death ($\langle\omega^i\rangle_1\simeq0$ and $\langle\omega^i\rangle_2\simeq0$) for the excitatory and inhibitory nodes even at the smallest value of $\lambda_+$. The values of $r^{link}_1$ and $r^{link}_2$ freeze to $0.99$ for any value of $\lambda_+$ implying the fact that there exist robust and distinct locally clustered excitatory and inhibitory phases in the respective incoherent regions. Moreover, the strong $\lambda_-$ gives rise to even steeper $r_1$ jump and more suppressed $r_2$ jump at further  value of $\lambda_+$ than that for the case of $\lambda_-=-0.8$.}

{Hence, it is evident that as $\lambda_-$ in the inhibitory layer strengthens, it shrinks width of the spread of effective frequencies of the excitatory and inhibitory nodes so as to form distinct local clusters of the excitatory and inhibitory phases. For a $\lambda_-$, when $\lambda_+$ becomes strong enough to overcome the suppressive effect of the inhibitory layer, the distinctly clustered excitatory phases abruptly get synchronized and give rise to ES transition. Nonetheless, $\lambda_+$ in the inhibitory layer can not win over compelling inhibition, therefore, the robust local clusters of inhibitory phases do not construct giant global cluster, in turn, ES transition get suppressed.} 

{Next, we investigate the impact of inhibitory coupling strength on hysteresis width of the excitatory layer in Fig.\ref{figure3_n1}. The increase in strength of $\lambda_-$ appears to widen the hysteresis width of forward and backward transitions of the excitatory layer.
In Fig.\ref{figure3_n2}, we also study the behavior of order parameter $r$ and $r^{link}$ of the entire multiplex network with increment in $\lambda_+$ for different values of $\lambda_-$. $r$ and $r^{link}$ for different strength of $\lambda_-$ follow the behavior of those of excitatory layer (see Fig. \ref{figure3}) except that the value of $r$ (of multiplex network) for any $\lambda_-$ drops approximately between $0.5-0.65$, which is perceptible as it represents the combined coherence of excitatory and inhibitory layers.}
\begin{figure}[t]
	\centering
	\includegraphics[height=4.5 cm,width=8.5 cm]{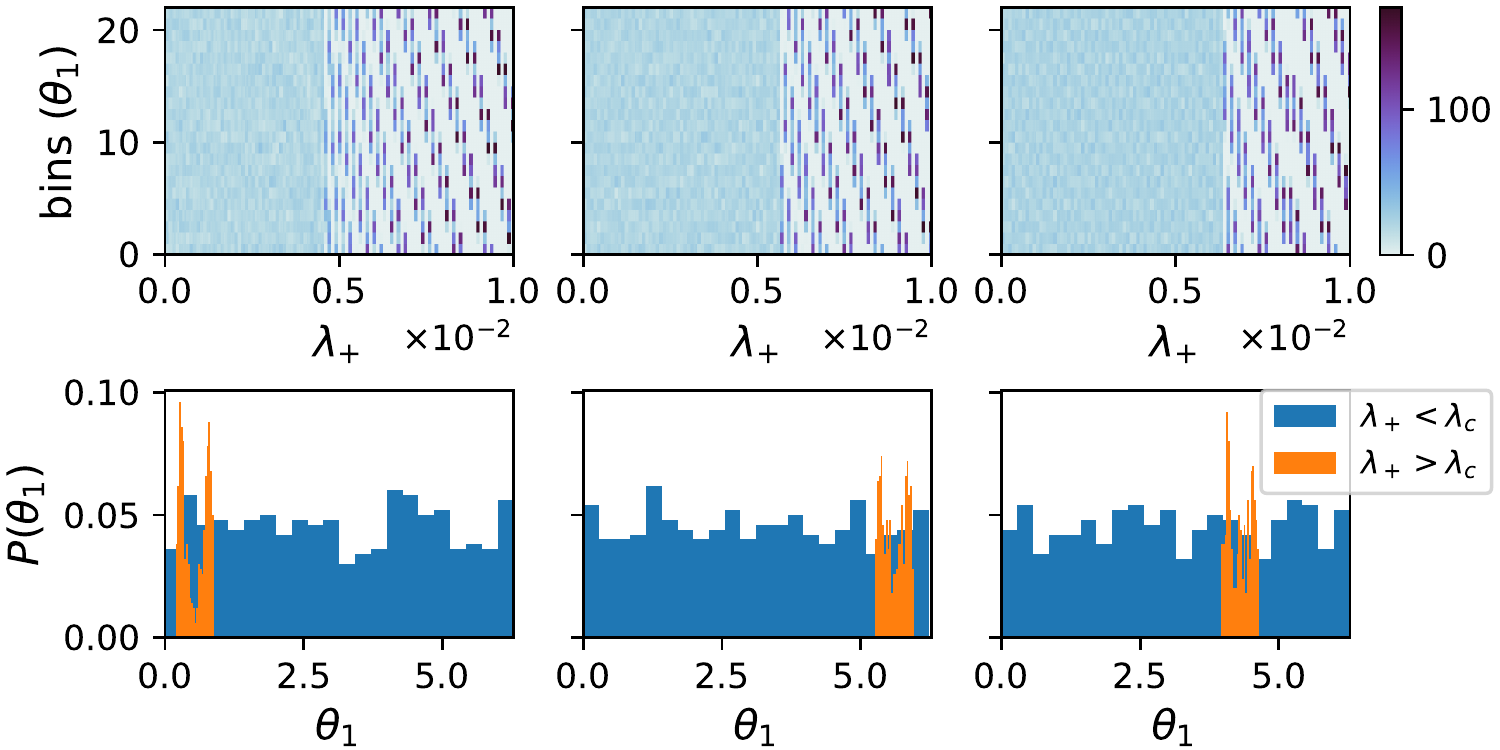}\\
	\includegraphics[height=4.5 cm,width=8.5 cm]{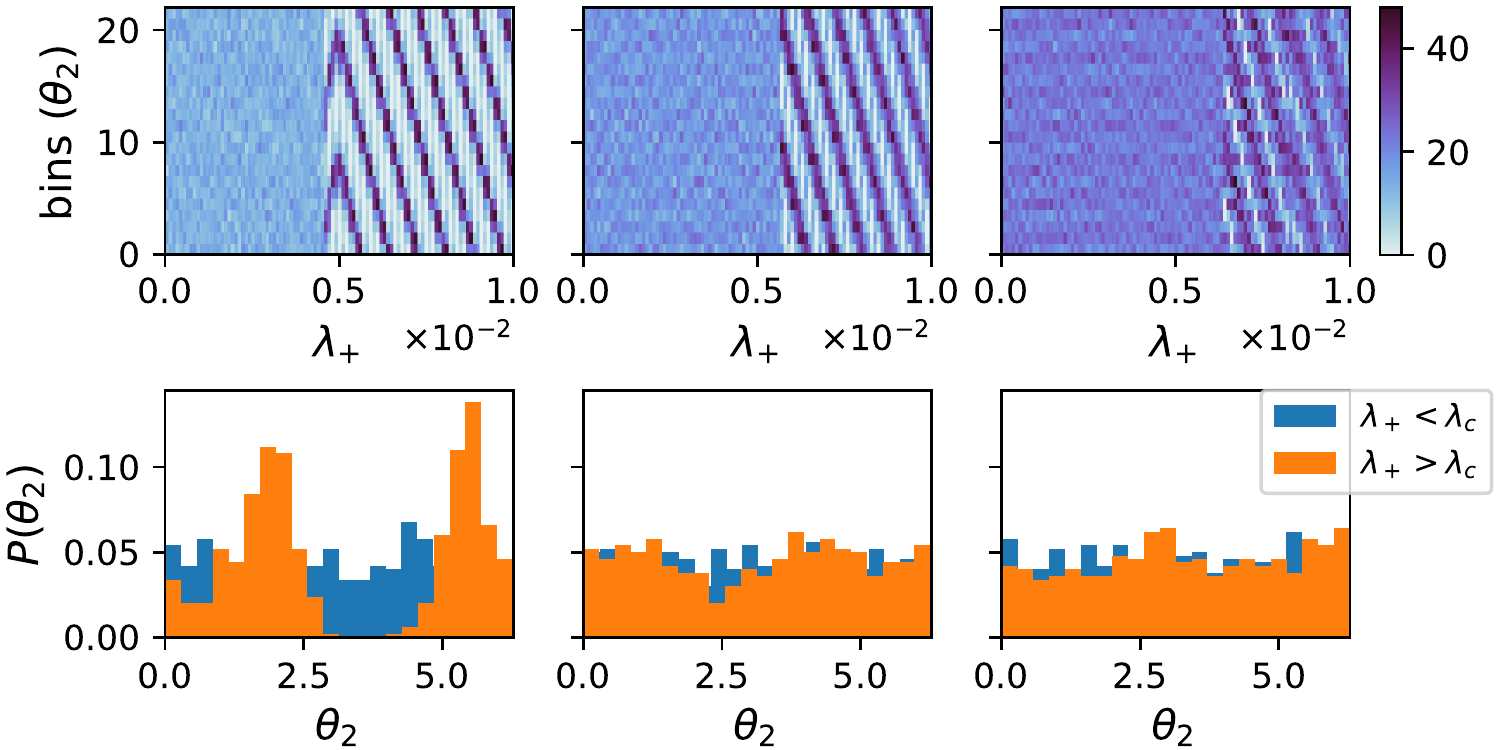}\\
	\caption{(Color Online) Binned excitatory $\theta_1$ (top row-panels) and inhibitory $\theta_2$ (3rd row-panels) phases with each increase in $\lambda_+$ associated with inhibitory coupling strength $\lambda_-=-0.5$ (left panels), $-1$ (middle panels) and $-2$ (right panels). 2nd and bottom row-panels depict distributions of the excitatory and inhibitory phases, respectively, before ($\lambda_+<\lambda_c$) and after ($\lambda_+>\lambda_c$) the onset of ES. Results are presented for $N=500$ nodes in each layer and $D_x=2$.}
	\label{figure3ab}
\end{figure}

\begin{figure}[t]
	\centering
	\includegraphics[height=4.5 cm,width=8.5 cm]{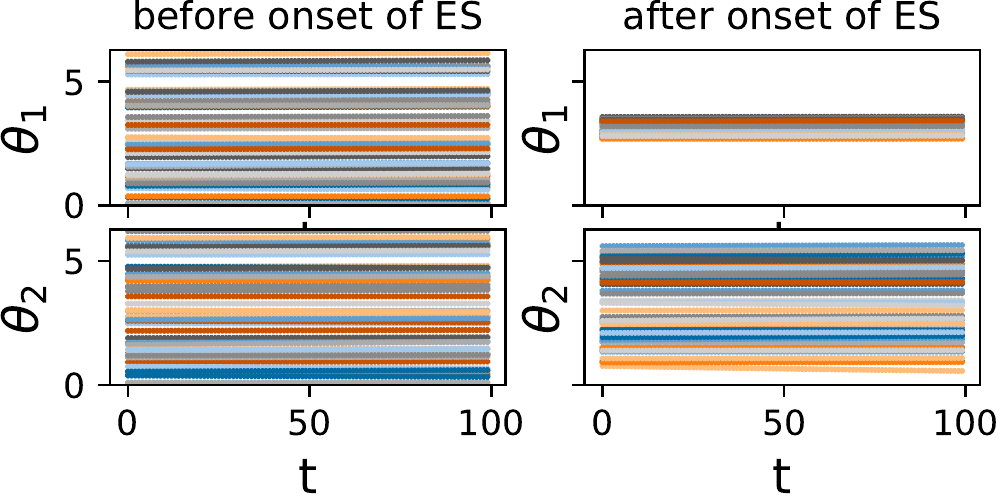}
	\caption{(Color Online) Time series of the phases of the nodes in the excitatory (top panels) and inhibitory (bottom panels) layers before (left panels) and after (right panels) onset of ES. Time series are obtained for $N=50$ nodes in each layer interacting under the effect of $\lambda_{-}=-0.5$ and $D_x=2$.}
	\label{figure3c}
\end{figure}

\begin{figure}[t]
	\centering
	\includegraphics[height=8 cm,width=8 cm]{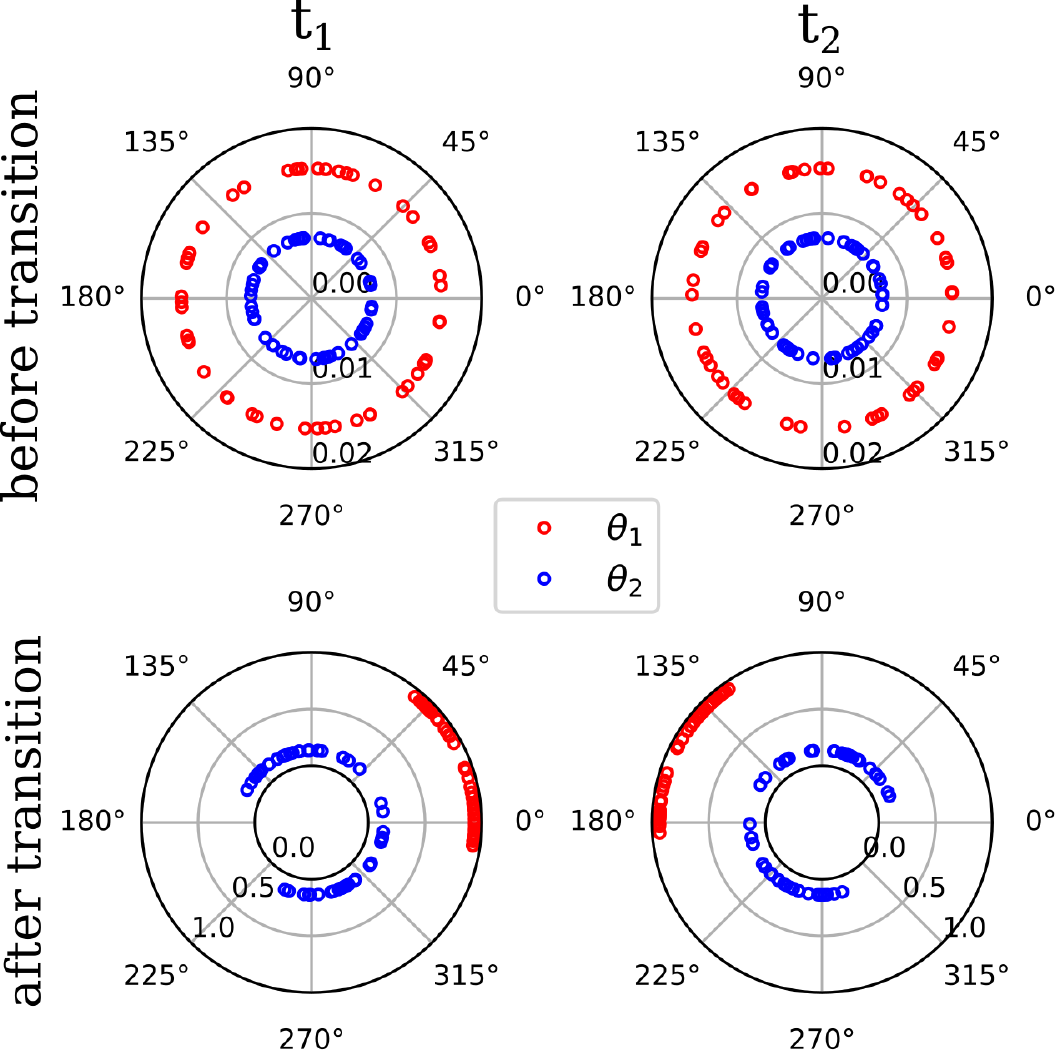} 
	\caption{(Color Online) Radar plots of the excitatory GC ($\theta_1$) and the inhibitory regular ring ($\theta_2$) phases at two different instants $t_1=1$ and $t_2=500$ before (top panels) and after (bottom panels) the triggering of ES. The radar plots are presented for $D_x=2$ and $N=50$ nodes in each layer with natural frequencies uniformly drawn in the range $0$ and $1$. All $\theta_1$ ($\theta_2$) rotate with constant frequency maintaining the angular distances with each other over the time.}
	\label{figure_3d}
\end{figure}

\textbf{\em Phase distribution $P(\theta)$ of each layer with $\lambda_-$}:
{In Fig.~\ref{figure3ab}, 1st and 3rd row-panels illustrate the binned phases of the excitatory and inhibitory layers, respectively, as a function of coupling strength $\lambda_+$ for different values of inhibitory coupling strength. 2nd and 4th row-panels depict distributions of the excitatory $\theta_1$ and inhibitory $\theta_2$ phases, respectively, for a value of $\lambda_+$ less than ($\lambda_+<\lambda_c$) and greater than ($\lambda_+>\lambda_c$) ES threshold $\lambda_c$. The top panels for $\theta_1$ make it apparent that before the onset of ES ($\lambda_+<\lambda_c$) for different values of $\lambda_-$, the excitatory phases are uniformly clustered into uniform bins covering the entire phase range [$0, 2\pi$], which becomes further apparent from histograms (blue) in corresponding 2nd row-panels. Also, after the onset of ES ($\lambda_+>\lambda_c$) for different values of $\lambda_-$, the excitatory phases are nearly synchronized and spread out over a few sequenced bins (top panels). The distribution for a $\lambda_+>\lambda_c$ (orange) in the 2nd row-panels further confirm phase synchronization with narrowly distributed bi-modal peaks.
In similar fashion, for any value of $\lambda_+$ in the incoherent region, the inhibitory phases $\theta_2$ are also uniformly distributed in the entire phase range [$0, 2\pi$] for different values of $\lambda_-$, which is also corroborated from distribution ($\lambda_+<\lambda_c$, blue) in corresponding 4th row-panels. However, post the outset of ES ($\lambda_+>\lambda_c$), $\theta_2$ are broadly distributed in bi-modal peaks for lower strength of $\lambda_-$. As the strength of $\lambda_-$ is increased further, $\theta_2$ now gradually start following uniform distribution covering the range [$0, 2\pi$], which is manifested from histograms (orange) in bottom panels.}

\textbf{\em Phase-evolution of each layer}:
{We also study time-series of the phases of the nodes in each layer before and after the onset of ES (see Fig~\ref{figure3c}). We find that after the onset of ES, the phases in the excitatory layer are synchronized, while the phases in the inhibitory layer remain stationary yet do not exhibit global synchrony as the strong inhibition within the layer suppresses it. Before the onset of ES, the inhibitory layer exhibits stationary phases because of the strong intralayer inhibition felt, nevertheless the excitatory layer also surprisingly exhibits stationary phases but due to inhibition felt through intralayer links.}

{Fig.~\ref{figure_3d} depicts the radar representation of phases of the excitatory (GC) and inhibitory (regular) layers at two different times $t_1$ and $t_2$ prior and post the transition. Here, the natural frequencies of $\theta_1$ and $\theta_2$ are uniformly selected between $0$ and $1$ instead of $-0.5$ and $0.5$ as considered for Fig~\ref{figure3c} where phases of both the layers appear stationary with time prior and post the transition. It is evident from the Fig.~\ref{figure_3d} that for both before and after transition cases, the phases $\theta_1$ ($\theta_2$) rotate with the same frequency and preserve the angular distance with each other and hence their distribution.}

\textbf{\em Theoretical insight}: {The employed technique for the emergence of ES works under the constraint that the inhibitory coupling strength must be more significant than the excitatory coupling strength, i.e., $\lambda_- >> \lambda_+$. Therefore, $\lambda_-$ is kept fixed to a value much larger than the critical value of $\lambda_+$ required for transition of the excitatory layer so that the excitatory nodes all the time remain under the impression of suppression. Under the effect of this constraint, each node in the inhibitory layer is subject to strong negative interactions from all intralayer neighbors and a single positive inter-layer interaction, thereby suppressing global synchronization in the inhibitory layer. The inhibitory nodes are driven by very strong inhibition to form distinct local clusters of uniformly distributed inhibitory phases leading to oscillation death (see Fig.~\ref{figure3ab} and bottom panels of Fig~\ref{figure3}). However, each node in the excitatory layer is subject to positive coupling with all its intralayer neighbors and strong negative coupling through the only inter-layer inhibitory node. Under the effect of constraint and appropriate multiplexing strength, the compelling suppression from each inhibitory node on its mirror excitatory node impedes coherence in the excitatory layer to some extent until a critical coupling $\lambda_+$ at which abrupt ES transition takes place. The excitatory nodes in the incoherent state following the mirror inhibitory nodes also tend to construct distinct local clusters of phases which under the significant inhibition lead to oscillation death (see Fig.~\ref{figure3ab} and top panels of Fig~\ref{figure3}). Thus, appropriate choices of the parameter $D_x$ and $\lambda_-$ enable us to control the characteristics of emergent ES.}

Next, we discuss how various dynamical and structural properties or parameters associated with the multiplex network under consideration affect the occurrence of ES in the excitatory layer.
\begin{figure}[t]
 \centering
  \includegraphics[height=4.5 cm, width= 8.5 cm]{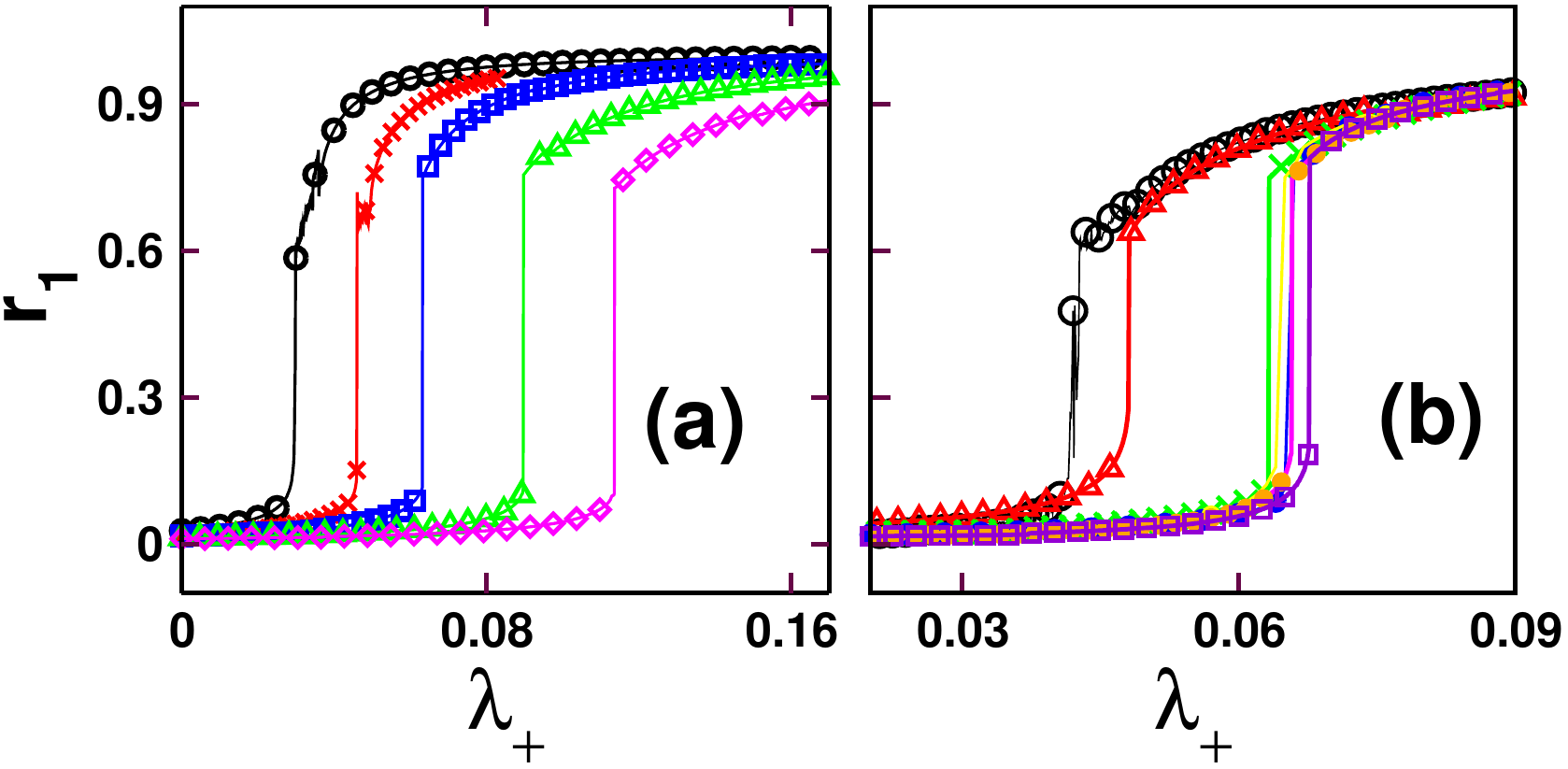}
    \caption{(Color Online) $r_1$ as a function of $\lambda_+$ for the multiplex network comprised of a positive GC and a negative regular networks for different values of (a) inter-layer coupling strength $D_x=1$ (circle), 1.5 (X), 2 (square), 3 (triangle up) and 4 (diamond), and (b) inhibitory coupling strength $\lambda_-=-0.5$ (black circle), -1 (red triangle up), -2 (green X), -3 (yellow circle), -4 (blue circle), -5 (orange circle) and -8 (violet square).}  
     \label{figure_4a}
\end{figure}

\textbf{\em Impact of $D_x$ on ES}: Here, we show that the inter-layer coupling strength $D_x$ plays a crucial role in determining the onset of ES, i.e., the critical coupling strength $\lambda_+^{c}$. 
Fig.\ref{figure_4a}(a) illustrates the ES transition in  the excitatory GC layer for different values of inter-layer coupling strength $D_x$. An apparent increase in $\lambda_+^c$ is observed with the increment in the inter-layer coupling strength. This observation can be attributed to the fact that an increment in $D_x$ leads to more impact of multiplexing with the inhibitory layer thereby leading to more suppression of synchrony in the excitatory layer. The strong influence of the negative coupling from the inhibitory layer through $D_x$ restrains the nodes of the excitatory layer from converging towards the synchronous state. However, at a sufficiently high coupling strength $\lambda_+^{c}$, all the nodes in the excitatory layer exhibits an abrupt jump and formation of the largest synchronized cluster takes place. Thus, $\lambda_+^{c}$ increases with each increase in $D_x$.
Furthermore, it presents the importance of choosing a multiplex framework where along with the negative coupling, the inter-layer coupling strength also contributes in determining the coupling strength threshold for the ES transition.

\textbf{\em Impact of $\lambda_-$ on ES:} Further, we investigate the impact of the strength of inhibitory coupling on ES observed in the excitatory layer. As the magnitude of the inhibitory coupling $\lambda_-$ in the inhibitory layer is increased, initially both the critical coupling $\lambda_+^{c}$ and the abrupt jump size in $r_1$ increase considerably (Fig.\ref{figure_4a}(b)). However, beyond a certain value of the inhibitory coupling, the increments in the value of both $\lambda_+^{c}$ and abrupt jump size slow down and start saturating to their respective constant values. Hence, the degree of suppression in the excitatory layer increases significantly up to a certain value of the inhibitory coupling strength and beyond which the degree of suppression gradually starts saturating.

\begin{figure}[t]
 \centering
  \includegraphics[width=0.6\columnwidth]{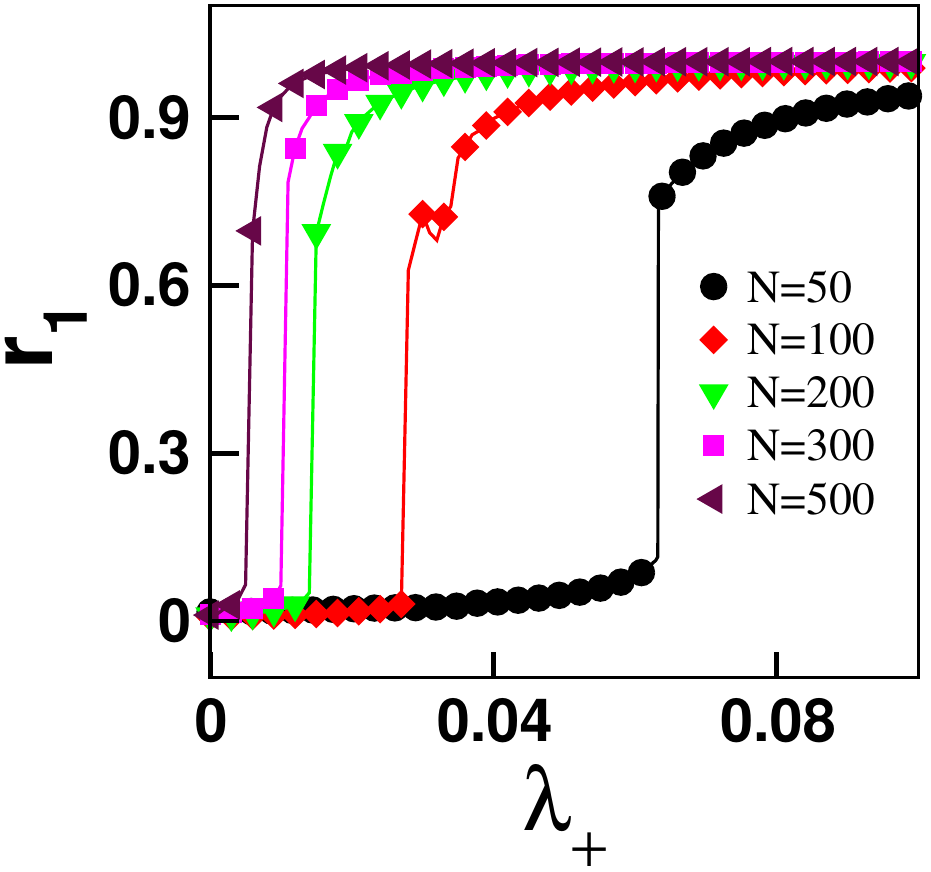}
    \caption{(Color Online) $r_1$ as a function of $\lambda_+$ for a multiplex network comprised of a positive GC and a negative regular network for different values of the network size.}  
     \label{figure_4b}
\end{figure}

\textbf{\em Impact of network size on ES}: We carry forward our numerical analysis for the duplex network, with different network sizes. Fig.\ref{figure_4b} depicts the effect of network size $N$ on ES transition observed in the excitatory GC layer. It is apparent that with increase in the size of network the critical coupling strength decreases as a large number of nodes accelerates contributing to the onset of the ES process.

So far, we have demonstrated that the inhibitory coupling in the regular layer accounts for the emergence of ES in the excitatory GC layer. However, it is important to validate the robustness of the existence of ES induced by the inhibitory coupling for different network architectures of the two layers.

\begin{figure}[t]
    \centering
    \includegraphics[width=1.0\columnwidth]{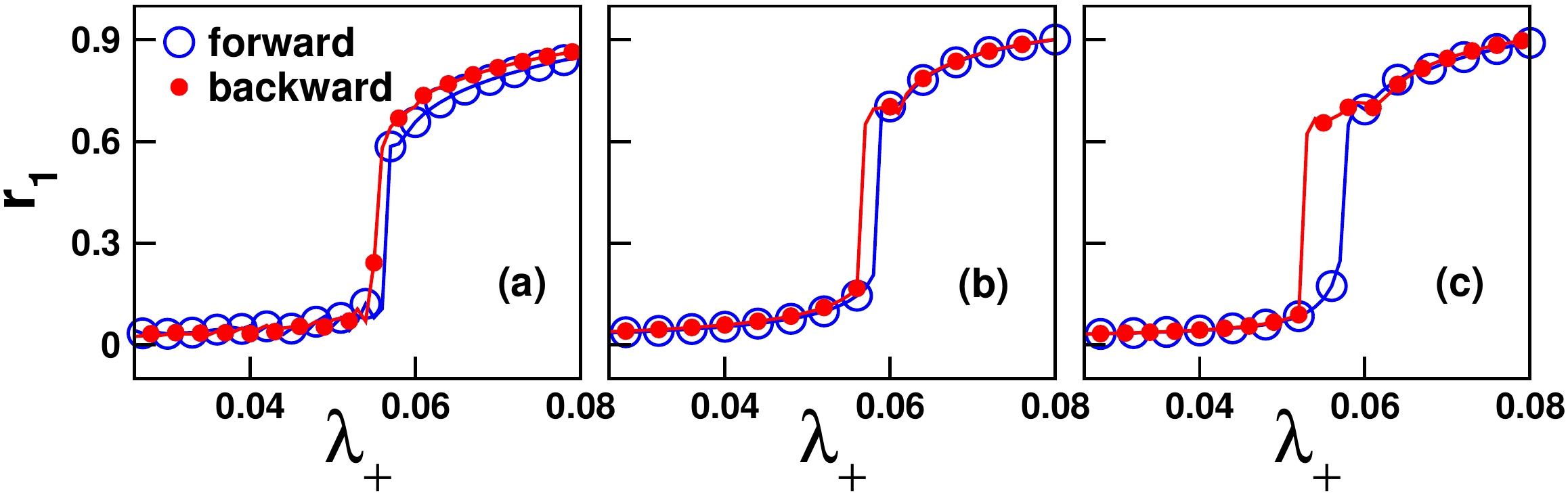}
    \caption{(Color Online) $r_1$ as a function of $\lambda_+$ when excitatory layer is fixed as GC network and it is multiplexed with different network architectures for the inhibitory layer, (a)  GC, (b) ER , and (c) SF. Here $\langle k_{1} \rangle=\langle k_{2} \rangle=10$ and $D_x=2$.}
    \label{figure5}
 \end{figure}

\textbf{\em Robustness of ES against network topology of the inhibitory layer}: Here we validate the emergence of ES by selecting a different network topology for the inhibitory layer while keeping topology of the excitatory layer fixed to the GC network. The excitatory layer exhibits ES transition to synchrony when the inhibitory layer is tested for GC, ER and SF topologies (Fig.\ref{figure5}).
Hence, we emphasize that a positively coupled layer when multiplexed with a negatively coupled layer of any network topology, gives rise to the ES transition. 
 
\textbf{\em Robustness of ES against network topology of the excitatory layer}:
Here we validate the emergence of ES against different network topologies chosen for the excitatory layer. For the validation of the occurrence of ES, we consider scale-free (SF), random ER or regular ring networks to represent the excitatory layer. We fix topology of the negative layer to a regular network.

\begin{figure}[t]
	\centering
	\includegraphics[width=1.0\columnwidth]{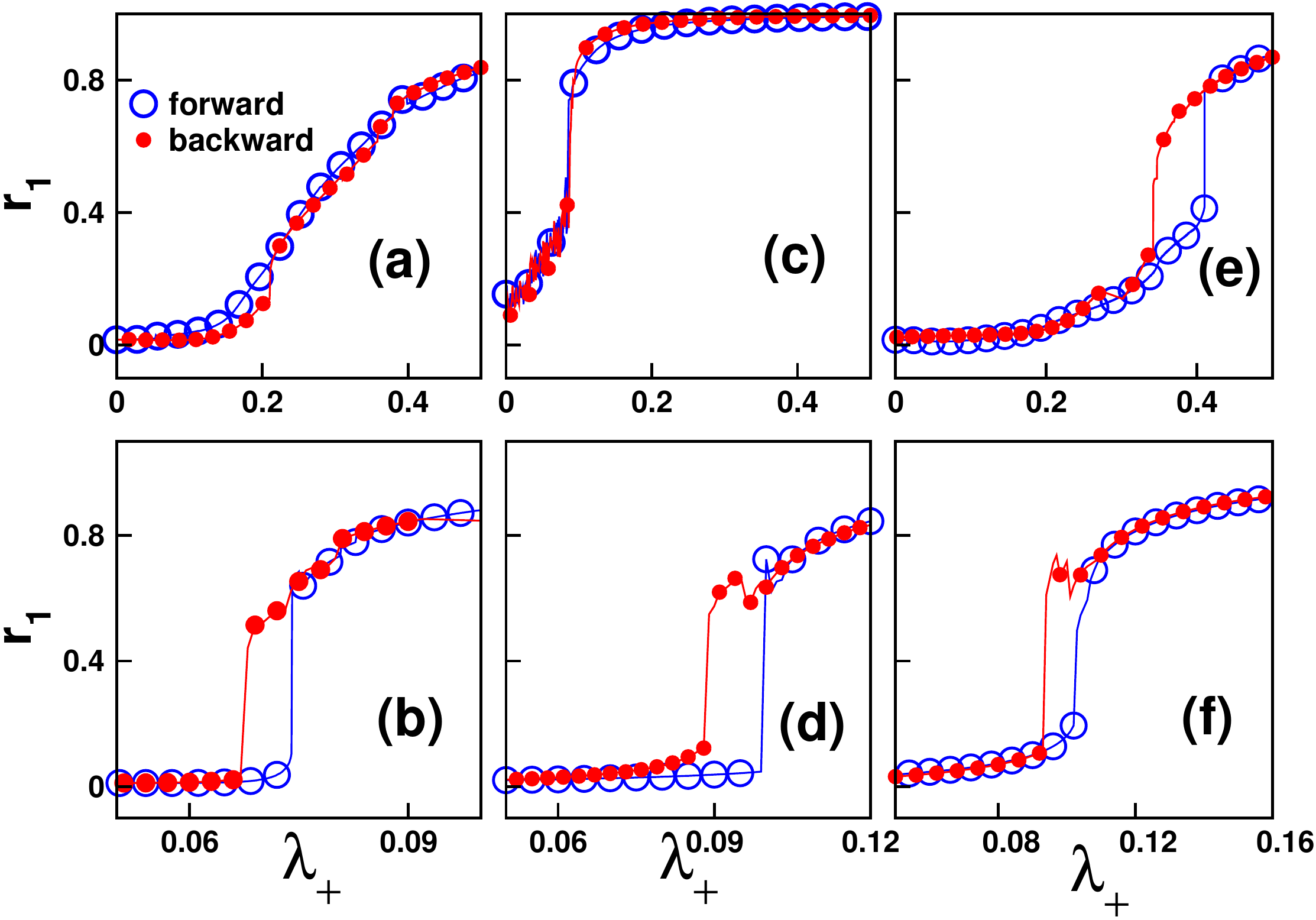}
	\caption{(Color Online) $r_1$ as a function of $\lambda_+$ for different network architecture of excitatory layer. For all the cases excitatory layer is multiplexed with an inhibitory regular layer, $\langle k_{2} \rangle=10$. (a) and (b) correspond to SF layer; (c) and (d) correspond to regular layer; and (e) and (f) correspond to the ER layer for $\langle k_{1} \rangle=10$ and $\langle k_{1} \rangle=30$, respectively.}
	\label{figure6}
\end{figure}

Fig.\ref{figure6}(a) depicts that upon multiplexing, a SF layer with $\langle k_{1} \rangle=10$ displays a second order transition, while a rather dense SF layer with $\langle k_{1} \rangle=30$ (Fig.\ref{figure6}(b)) does show an ES transition.
It is known that an isolated sparse SF network requires higher coupling strength to get synchronized as compared to an isolated ER network of the same connectivity, whereas a dense SF network gets synchronized at relatively weak coupling strength. This is the reason SF layer with $\langle k_{1} \rangle=10$ does not show ES, however SF layer with $\langle k_{1} \rangle=30$ exhibits ES upon being multiplexed with the inhibitory layer with $\lambda_-=-2$. 

Fig.\ref{figure6}(c) and (d) correspond to the case when the excitatory layer is a regular ring network. In this case, too, a sparse regular network with $\langle k_{1} \rangle=10$ lead to a second-order transition, whereas a rather dense regular network with $\langle k_{1} \rangle=30$ gives rise to an ES transition. It is known that an isolated regular network does not synchronize if the node degree is very small. Next, in Fig.\ref{figure6}(e), a random ER layer even with $\langle k_{1} \rangle=10$ exhibits an ES transition with hysteresis loop. This can be attributed to the fact that an isolated ER network gets synchronized at lower coupling strength as compared to an isolated SF network of the same connectivity. However, a stronger average connectivity $\langle k_{1} \rangle=30$ of ER leads to a stronger ES transition (see Fig.\ref{figure6}(f)). 
The comparisons carried out against a variety of network topologies reveal that the inhibitory layer can induce an ES transition in the excitatory layer of any network topology provided the fact that it should be capable of achieving the synchronization at a relatively lower coupling strength in its isolation.

\section{Analytical insight}
	The time-evolution of Kuramoto oscillators in a duplex network given in \ref{eq:k_model_multi} can be rewritten in the following rearranged composite form
	\begin{equation}\label{eq:km_comp}
	\dot{\theta_i} = \omega_i + \lambda_{\pm}\sum^{2N}_{j=1}A_{ij}\sin(\theta_j-\theta_i)+(D_x-\lambda_{\pm})\sin(\theta_l-\theta_i),
	\end{equation}
	where $i=1,2,\cdots,2N$ and
	\[
	\lambda_{\pm}= 
	\begin{cases}
	\lambda_{+}, & \text{if } i\leq N\\
	\lambda_{-}, & \text{if } i>N
	\end{cases}
	\]
	and
	\[
	\theta_l= 
	\begin{cases}
	\theta_{i+N}, & \text{if } i\leq N\\
	\theta_{i-N}, & \text{if } i> N
	\end{cases}
	\]
	
	The local mean-field order parameter $r_i$ for a node $i$ having local average phase $\psi_i$ arising from $k_i$ neighbors is defined as
	\begin{equation}\label{eq:local_op}
	r_ie^{i\psi_i} = \frac{1}{(k_i+1)}\sum_{j=1}^{2N}A_{ij}e^{i\theta_j}.
	\end{equation}
	
	Eq.\ref{eq:km_comp} can be rewritten in terms of local order parameter $r_i$ as
	\begin{equation}
	\dot{\theta_i} = \omega_i + \lambda_{\pm}r_i(k_i+1)\sin(\psi_i-\theta_i)+(D_x-\lambda_{\pm})\sin(\theta_l-\theta_i),
	\end{equation}
	which can be simplified further in the following
	\begin{equation}\label{eq:km_finalEq}
	\dot{\theta_i} = \omega_i + \acute{\lambda_{\pm}}\sin(\psi_i-\theta_i+\alpha_{\pm}), \\
	\end{equation}
	where parameters $\alpha_{\pm}$ and $\acute{\lambda_{\pm}}$ are respectively defined as
	\begin{align}\label{eq:km_alpha}
	\tan\alpha_{\pm} & = \pm \frac{(D_x-\lambda_{\pm})}{\lambda_{\pm} r_i (k_i+1)} \quad \mathrm{and} \nonumber \\
	{\acute{\lambda_{\pm}}}^2 = & (\lambda_{\pm}r_i (k_i+1))^{2} + (D_x-\lambda_{\pm})^{2}.
	\end{align} 
	
\begin{figure}[t]
	\centering
	\includegraphics[width=1.0\columnwidth]{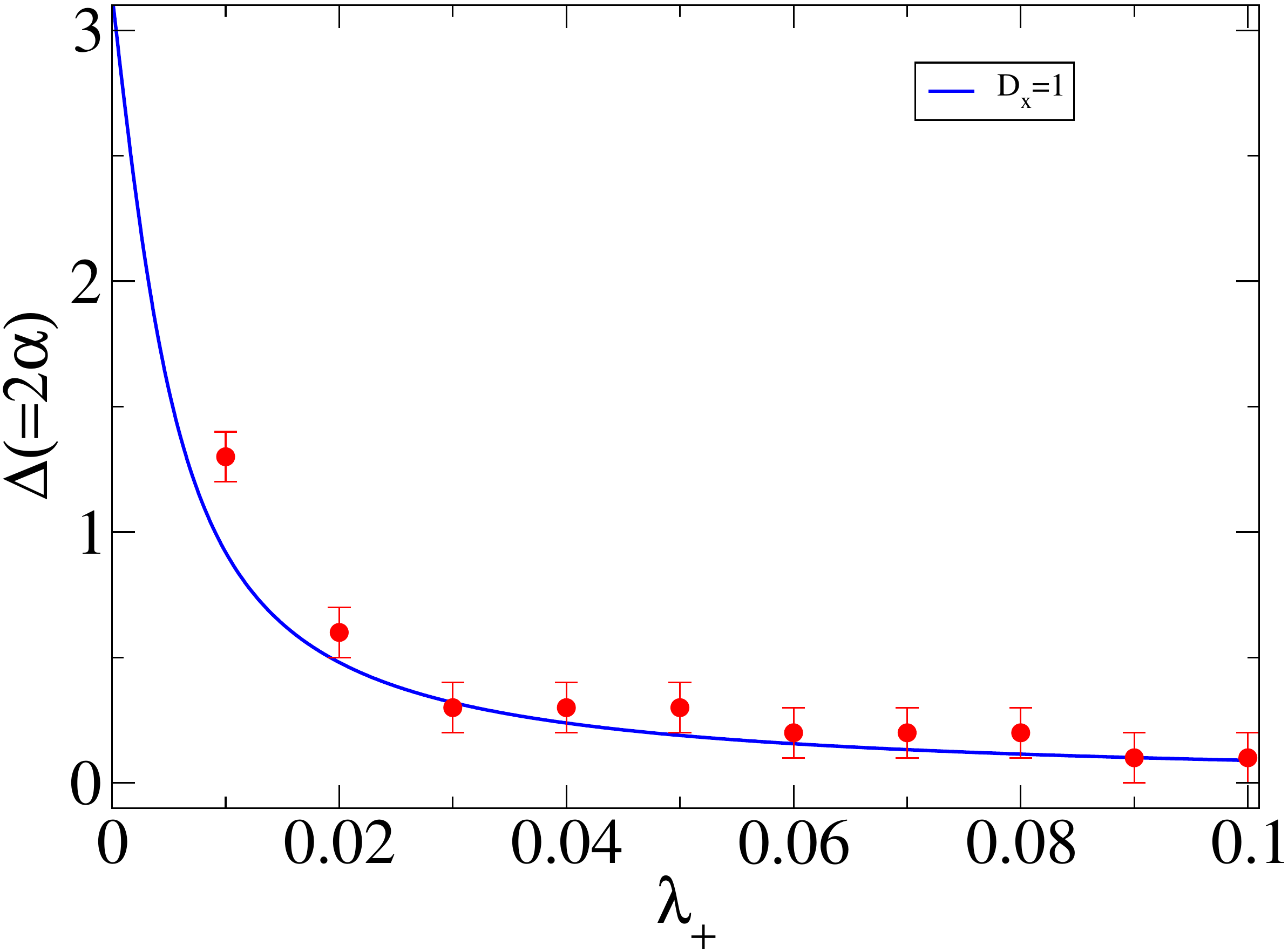}  
	\caption{(Color Online) Separation $\Delta$ \ref{eq:delta} between the synchronized bi-modal peaks of the excitatory phases as a function of $\lambda_+$. The blue line depicts the analytical prediction while red circle represents numerical estimation. The parameters are $N=200$, $D_x=1$ and $\lambda_-=-2$.} 
	\label{figure7} 
\end{figure}

	Eqs.~\ref{eq:km_finalEq} and \ref{eq:km_alpha} reveal that the nodes in both the layers have their local mean-field phases split into two new phases ($\psi_i\pm\alpha_{\pm}$) with total separation between them being 
\begin{equation}\label{eq:delta}
	\Delta = 2~ tan^{-1} \frac{(D_x-\lambda_{\pm})}{\lambda_{\pm} r_i (k_i+1)}.
\end{equation}
	Hence, the separation between the phases depend upon the strength of $D_x$ and $\lambda_{\pm}$. The splitting of phases can be evaded for the excitatory layer when $\lambda_{+}$ tends to $D_x$, i.e., $D_x\simeq\lambda_{+}$, but the same is not true for the inhibitory phases as inhibitory coupling strength is always smaller than intra-layer coupling strength ($\lambda_{-}< 0\leq D_x$).
	Fig.\ref{figure7} depicts $\Delta$, the separation between two synchronized sharp bi-modal peaks (as shown in Fig.\ref{figure3ab}), as a function of $\lambda_+$, corresponding to phases of the excitatory nodes. Hence, both numerical and analytical estimations match closely demonstrating validity of our results.
	
	Also, $\theta_l$ and $\psi_i$ obey the relation 
	$\tan\theta_l \tan\psi_i = -1$, which yields 
	\begin{equation}
	\theta_l = \psi_i \pm \frac{\pi}{2}+2n\pi, \quad  \forall n=0,1,2,\cdots. 
	\end{equation}
	This implies that the inhibitory nodes and their excitatory counterparts maintain a phase difference between of $\frac{\pi}{2}$, i.e., (($\psi_i-\frac{\pi}{2}$), ($\psi_i + \frac{\pi}{2}$), $n=0$). 

\section{Conclusion} 
{Earlier reported prerequisites for the emergence of ES in all the layers of a multiplex network require that at least one layer, in the absence of multiplexing, must be exhibiting ES to trigger off ES in all the multiplexed layers.
In the present study, strikingly, we have shown that one can induce ES in a layer by multiplexing it with an inhibitory layer, hence evading the precondition of having a layer already showing ES in its isolation.
It is revealed that the inclusion of a layer with all negatively coupled links impedes the formation of the largest synchronized cluster in the excitatory layer by propagating suppression via inter-layer links, in turn, resulting in occurrence of ES. It is further shown that such emergence of ES originating from inhibitory coupling remains true for a variety of combinations of network topologies selected for the multiplexed layers. Also, the scheme employed provides us control over the induced ES transition by tuning structural parameters such as average degrees of the layers, inter- or intra-layer coupling strengths. Hence, in the present investigation, we have successfully devised a technique to achieve ES by incorporating inhibition through a single layer in a multiplex framework.}

The results presented here have application in understanding synchronization in those systems which in addition of having inherent multiplex architecture have negative coupling between inter units. For example two-layered epithelial-mesenchymal transition process, in which growth involves rapid proliferation of epithelial cells in one layer, while another layer of mesenchymal cells type suppresses proliferation of epithelial cells. In the hair cycle (whole mouse skin), mRNA gene periodic expressions of these two types of cells play a crucial role in the dynamics of growth of hair cells \cite{haircycle}. The positively and negatively coupled layers of gene clusters associated with rapidly growing epithelial cells and inhibitory mesenchymal populations in the hair follicles, respectively, are responsible for the hair cycle dynamical process. In another example, thymic mesenchymal cells derived retinoic acid regulates epithelial cells development in embryonic thymus \cite{thymus}.
Another example of multiplex network consists of positive and negative coupling is in ecological systems in which ecological balance exists between facilitation (represents positive interaction) and competition (represents negative interaction) in plant communities. These interactions play a key role in the structure and organization of plant communities \cite{ecological_example}. A similar example is the one that of opinion formation in the case of public polling \cite{poll_example}. Those people whose opinion match and they agree, share positive interaction while those who disagree experience negative interactions, a complex interaction between positive and negative coupling determines the outcome. 

Hence, our investigation about understanding the role of inhibition or suppressive effect on dynamics and its regulation in the multiplex framework can be constructive in learning the underlying dynamics of regulatory biological systems, ecological systems and formation of a common opinion among a group of people.\\

\begin{acknowledgments} 
SJ acknowledges Govt of India, DST grant EMR/2016/001921, BRNS grant 37(3)/14/11/2018-BRNS/37131 and CSIR grant 25(0293)/18/EMR-II for financial support. VR is thankful to Govt of India, DST grant DST/INSPIRE Fellowship/[IF180308]. ADK acknowledges Govt of India CSIR grant 25(0293)/18/EMR-II for RA fellowship. AY thanks to Govt of India DST grant EMR/2016/001921 for SRF fellowship.
\end{acknowledgments}

\end{document}